\documentclass[journal,doublecolumn,10pt]{IEEEtran}

\usepackage{epsfig,latexsym}
\usepackage{float}
\usepackage{indentfirst}
\usepackage{amsmath}
\usepackage{amssymb}
\usepackage{times}
\usepackage{subfigure}
\usepackage{psfrag}
\usepackage{cite}
\usepackage{lastpage}
\usepackage{fancyhdr}
\usepackage{color}

\begin{document}%
\title{ {\huge On the Performance of Non-Orthogonal Multiple Access in 5G Systems with Randomly Deployed Users}}

\author{ Zhiguo Ding, \IEEEmembership{Member, IEEE}, Zheng Yang, Pingzhi Fan, \IEEEmembership{Senior Member, IEEE},
  and  H. Vincent Poor, \IEEEmembership{Fellow, IEEE}\thanks{
Z. Ding and H. V. Poor are with the Department of
Electrical Engineering, Princeton University, Princeton, NJ 08544,
USA.   Z. Ding and Z. Yang  are with the School of
Electrical, Electronic, and Computer Engineering, Newcastle
University, NE1 7RU, UK. Z. Yang and P. Fan are with the Institute of Mobile Communications, Southwest Jiaotong University, Chengdu, China.  }\vspace{-1em}} \maketitle
\begin{abstract}
 In this letter, the performance of non-orthogonal multiple access (NOMA) is investigated in a cellular downlink scenario  with randomly deployed users. The developed analytical results show that NOMA can achieve superior performance in terms of ergodic sum rates; however, the outage performance of   NOMA     depends critically  on the choices of the users' targeted data rates and allocated power. In particular,  a wrong choice of the targeted data rates and allocated power can lead to a situation in which  the user's outage probability  is always one, i.e. the user's targeted quality of service  will never be  met.
\end{abstract}
\vspace{-1.5em}
\section{Introduction}
 4G mobile networks, such as long term evolution (LTE), are being deployed worldwide, and   research to define  the next generation mobile network is now receiving   considerable  attention \cite{Li5G}. Particularly non-orthogonal multiple access (NOMA) has been recognized as a promising multiple access techniques for     5G   networks  due to its superior spectral efficiency \cite{NOMAPIMRC}.

In this letter, the performance of NOMA is investigated in a downlink network with   randomly deployed mobile users. In particular,  the performance of NOMA is evaluated in two types of situations. {\it Firstly} we consider the case in which  each user has a targeted data rate that  is determined by its quality of service (QoS). In this situation, the outage probability is an ideal metric  for performance evaluation since it measures the ability of NOMA to meet the users' QoS requirements. The developed analytical results demonstrate that the choices of the users' targeted data rates and allocated power are critical to their outage performance. In particular,  there is a critical condition about   these system parameters, such that  the outage probability is always one if the condition is not satisfied. But provided that this condition is satisfied, NOMA can ensure that the multiple users access the shared wireless medium and  experience  {\it the same diversity order} as  conventional orthogonal multiple access (MA) techniques.

{\it Secondly} we consider the case in which users' rates are allocated opportunistically according to their channel conditions. In this situation, the ergodic sum rate achieved by NOMA is studied. Particularly the high signal-to-noise ratio (SNR) approximation of the ergodic rate is developed first, and an asymptotic study of the sum rate is carried out by focusing on the case in which  the number of mobile users goes to infinity. The provided analytical and simulation  results demonstrate that NOMA can achieve  superior performance in terms of ergodic sum rates. For example, the more users that join in cooperation, the larger sum rate the NOMA can achieve, which demonstrates that  NOMA is spectrally efficient. In addition asymptotic studies show that NOMA can approach an upper bound on the multi-user system which is achieved by  always allocating all the  bandwidth resources to the user with the best channel condition.
\vspace{-0.5em}

\section{NOMA Transmission Protocol}
Consider a cellular downlink transmission scenario, in which the base station is located at the center of a disc, denoted by $\mathcal{D}$,  with   radius $\mathcal{R}_D$, and   $M$ users are uniformly distributed within the disc. The channel between the $m$-th user and the base station is denoted by $h_m$, and $h_m=\frac{\tilde{g}_m}{\sqrt{1+d_m^{\alpha}}}$, where $\tilde{g}_m$ denotes the Rayleigh fading channel gain, $\alpha$ is the path loss factor,  and $d_m$ denotes the distance from the user to the base station.
Without loss of generality,  the channels are sorted as $|h_{1}|^2\leq \cdots \leq |h_{M}|^2$.
According to the NOMA protocol, the base station will send $\sum^{M}_{m=1}\sqrt{a_mP}s_m$, where $s_m$ is the message for the $m$-th user, $P$ is the transmission power, and $a_m$ is the power allocation coefficient, i.e.  $a_1\geq \cdots \geq a_M$. Therefore the observation at the $m$-th user is given by
\begin{eqnarray}
y_m = h_{m} \sum^{M}_{i=1}\sqrt{a_iP}s_i +n_m,
\end{eqnarray}
where $n_m$ denotes   additive noise. Successive interference cancellation (SIC) will be carried out at the users. Therefore the $m$-th user will   detect the $i$-th   user's message, $i<m$, and then remove the message from its observation, in a successive manner. The message for the $i$-th user, $i>m$, will be treated as noise at the $m$-th user.  As a result, the data rate achievable to the $m$-th user, $1\leq m \leq (M-1)$, is given by
\begin{eqnarray}
R_{m}= \log \left(1 + \frac{\rho|h_m|^2a_m}{\rho|h_m|^2\sum^{M}_{i=m+1}a_i+1} \right),
\end{eqnarray}
 conditioned on $R_{j\rightarrow m}\geq \tilde{R}_{j}$,   where $\rho$ denotes the transmit SNR, $\tilde{R}_j$ denotes the targeted data rate of the $j$-th user,  and $R_{j\rightarrow m}$ denotes the rate for the $m$-th user to detect the $j$-th user's message, $j\leq m$, i.e. $R_{j\rightarrow m}= \log \left(1 + \frac{\rho|h_m|^2a_j}{\rho|h_m|^2\sum^{M}_{i=j+1}a_i+1} \right)$. Note that the rate at the $M$-th user is  $R_M = \log(1+\rho |h_M|^2a_M)$.

In this letter two types of   $\tilde{R}_m$ are considered.
\subsubsection{Case I}  $\tilde{R}_m$ is determined by the users' QoS requirements, i.e. each user has a preset $\tilde{R}_m$.  In this case, it is important to examine the probability of the following two events. One is that  a user can cancel others users' messages,  i.e. $R_{j\rightarrow m}\geq \tilde{R}_{j}$, $j<m$, and the other is that NOMA can ensure the user's QoS requirements  to be satisfied, i.e. $R_{m}\geq \tilde{R}_{m}$. When both constraints are satisfied, the sum rate of NOMA is simply $\sum_{m=1}^{M}\tilde{R}_m$. Therefore the sum rate will not be of interest in this case, and it is important to calculate the  probabilities of the two events, as  shown in Section \ref{section outtage}.

\subsubsection{Case II} $\tilde{R}_m$ is determined opportunistically  by the user's channel condition, i.e.    $\tilde{R}_j=R_j$. Therefore it can be easily verified that the condition $R_{j\rightarrow m}\geq \tilde{R}_{j}$   always holds    since $|h_m|^2\geq |h_j|^2$ for $m>j$.
Consequently the sum rate achieved by NOMA is given by
 \begin{align}\nonumber
R_{sum} = \sum^{M-1}_{m=1}\log (1 + \frac{\rho|h_m|^2a_m}{\rho|h_m|^2\tilde{a}_m+1}  ) + \log(1+\rho|h_M|^2a_M),
\end{align}
where $\tilde{a}_m= \sum^{M}_{i=m+1}a_i$.
Therefore,  it is important to find the ergodic sum rate achieved by NOMA, as shown in Section \ref{section rate}.

\section{Density Functions of Channel Gains}
The evaluation of the outage probability and ergodic rates requires the density functions of the channel gains. Denote by $\tilde{h}$   an  {\it unordered} channel gain. Conditioned on the fact that the users are uniformly located in the disc,  $\mathcal{D}$, and small scale fading is Rayleigh distributed, the  cumulative density function (CDF), of the  unordered channel gain, $|\tilde{h}|^2$, is  given by \cite{Dingpoor132}
\begin{eqnarray}\label{cdf1}
F_{|\tilde{h}|^2}(y) = \frac{2}{\mathcal{R}_D^2}\int^{\mathcal{R}_D}_{0} \left(1-e^{-(1+z^\alpha)y}\right)zdz.
\end{eqnarray}
When $\alpha=2$, the above integral can be easily calculated as shown in \cite{Dingpoor132}. But for other choices of $\alpha$, the evaluation of the above integral is difficult, which  makes it challenging to carry out  insightful analysis.   In the following the Gaussian-Chebyshev quadrature will be used to find an approximation for the above integral \cite{Gaussiandd}. Rewrite \eqref{cdf1} as follows:
\begin{eqnarray} \nonumber
F_{|\tilde{h}|^2}(y) = \frac{1}{2}\int^{1}_{-1} \left(1-e^{-(1+(\frac{\mathcal{R}_D}{2}x+\frac{\mathcal{R}_D}{2})^\alpha)y}\right)\left( x+1\right)dx.
\end{eqnarray}
By applying   Gaussian-Chebyshev quadrature, we obtain the following simplified expression:
\begin{eqnarray}\label{cdf11}
F_{|\tilde{h}|^2}(y) \approx \frac{1}{\mathcal{R}_D} \sum^{N}_{n=1}w_n g(\theta_n),
\end{eqnarray}
where $g(x)= \sqrt{1-x^2}\left(1-e^{-c_ny}\right)
\left(\frac{\mathcal{R}_D}{2}x+\frac{\mathcal{R}_D}{2}\right)$, $N$ is a parameter to ensure a complexity-accuracy tradeoff, $c_n=1+\left(\frac{\mathcal{R}_D}{2}\theta_n+\frac{\mathcal{R}_D}{2}\right)^\alpha$, $w_n= \frac{\pi}{N}$, and $\theta_n =\cos \left( \frac{2n-1}{2N} \pi\right)$.

Consequently the probability density function (pdf) of the {unordered} channel gain can be approximated as follows:
\begin{eqnarray}\label{pdf1}
f_{|\tilde{h}|^2}(y) \approx \frac{1}{\mathcal{R}_D} \sum^{N}_{n=1} \beta_n e^{-c_ny}
,
\end{eqnarray}
where $\beta_n=w_n\sqrt{1-\theta_n^2}\left(\frac{\mathcal{R}_D}{2}\theta_n+\frac{\mathcal{R}_D}{2}\right)
c_n
$.  Compared to the original form in \eqref{cdf1}, the ones shown in \eqref{cdf11} and \eqref{pdf1}  can be used to  simplify the performance analysis significantly since they are linear combinations of exponential functions.

\section{Case I: Outage Performance of NOMA}\label{section outtage}
The outage events at the $m$-th user can be defined as follows. First define $E_{m,j}\triangleq\{R_{j\rightarrow m}< \tilde{R}_{j}\}$ as the event that the $m$-th user cannot detect the $j$-th user's message, $1\leq j\leq m$, and $E_{m,j}^c$ as the complementary set of $E_{m,j}$. The outage probability at the $m$-th user can be expressed as follows:
\begin{eqnarray}
\mathrm{P}^{out}_{m} = 1 - \mathrm{P}(E^c_{m,1}\cap \cdots \cap E^c_{m,m}).
\end{eqnarray}

The event $E^c_{M,M}$ is defined as
$E^c_{M,M} =\left\{    \rho|h_M|^2a_M  >\phi_M\right\}
$, and the other event $E^c_{m,j}$, $1\leq j\leq m$,  can be expressed as follows:
\begin{eqnarray}
E^c_{m,j} &=&\left\{   \frac{\rho|h_m|^2a_j}{\rho|h_m|^2\sum^{M}_{i=j+1}a_i+1}  >\phi_j\right\}
\\\nonumber &\underset{a}{=}&\left\{    \rho|h_m|^2\left(a_j - \phi_j \sum^{M}_{i=j+1}a_i\right)  >\phi_j\right\},
\end{eqnarray}
where $\phi_j=2^{\tilde{R}_j}-1$.  Note that the step (a) is obtained by assuming the following condition holds:
\begin{align}\label{condition}
 a_j  >\phi_j \sum^{M}_{i=j+1}a_i.
 \end{align}

Furthermore define $\psi_{j}\triangleq \frac{\phi_j}{\rho \left(a_j - \phi_j \sum^{M}_{i=j+1}a_i\right) }$ for $j<M$, $\psi_{M}\triangleq \frac{\phi_M}{\rho  a_M   }$, and  $\psi_m^*=\max\{\psi_{1}, \ldots, \psi_{m}\}$. As a result, the outage probability can now be expressed as follows:
 \begin{align}\label{outage 1}
&\mathrm{P}^{out}_{m} = 1 - \mathrm{P}( |h_m|^2>\psi_m^* )\\
\nonumber &=\int^{\psi_m^*}_{0}
\frac{M!\left(F_{|\tilde{h}|^2}(x)\right)^{m-1}\left(1-F_{|\tilde{h}|^2}(x)\right)^{M-m}f_{|\tilde{h}|^2}(x)}{(m-1)!(M-m)!}
dx,
\end{align}
which is obtained by analyzing  order statistics  \cite{David03}.

Note that when  $y\rightarrow 0$, the CDF  of the unordered channel gains can be approximated as follows:
\begin{eqnarray}
F_{|\tilde{h}|^2}(y) \approx \frac{1}{\mathcal{R}_D}
\sum^{N}_{n=1}\beta_n y,
\end{eqnarray}
and the approximation of the pdf is given by
\begin{eqnarray}
f_{|\tilde{h}|^2}(y) \approx \frac{1}{\mathcal{R}_D}
\sum^{N}_{n=1} \beta_n
\left(1-c_ny\right)
.
\end{eqnarray}
When $\rho\rightarrow \infty$, $\psi^*_m\rightarrow 0$. Therefore a high SNR approximation of  the outage probability is given by
\begin{align}\nonumber
\mathrm{P}^{out}_{m}  &\approx \tau_m\int^{\psi_m^*}_{0} \left(\eta  x\right)^{m-1}\left(1-\eta x\right)^{M-m}\frac{1}{\mathcal{R}_D} \\ &\times \sum^{N}_{n=1} \beta_n \left(1-c_nx\right)  dx\approx \frac{\tau_m}{m}\eta^{m}  \left(\psi_m^*\right)^m,\label{outage 2}
\end{align}
where $\eta=\frac{1}{\mathcal{R}_D} \sum^{N}_{n=1}\beta_n $ and $\tau_m=\frac{M!}{(m-1)!(M-m)!}$. Therefore the diversity order achieved by NOMA is given by
\begin{align}\label{outage 3}
\mathrm{P}^{out}_{m}\rightarrow \frac{1}{\rho^m}.
\end{align}
The result in \eqref{outage 3} demonstrates that the $m$-th user will experience a diversity order of $m$. This is better than a conventional orthogonal  MA scheme with a randomly scheduled user whose diversity order is one. Compared to opporunitstic user scheduling,  NOMA will achieve better spectral efficiency and user fairness  since all the users are served at the same time, frequency and spreading code.

It is worthy to  point out that the superior outage performance achieved by NOMA is conditioned on the constraint in \eqref{condition}. When such a condition is not satisfied, e.g. $ a_j  \leq \phi_j \sum^{M}_{i=j+1}a_i$, the user's outage probability is always one, i.e. $\mathrm{P}^{out}_{m}=1$, as shown in Section \ref{section numerical}.

 \section{Case II: Ergodic Sum Rate of NOMA}\label{section rate}
When $\tilde{R}_j=R_j$, the ergodic sum rate is given by
\begin{eqnarray}\nonumber
R_{ave}& =& \sum^{M-1}_{m=1}\int^{\infty}_{0}\log \left(1 + \frac{x\rho a_m}{x\rho \tilde{a}_m+1} \right)  f_{|h_m|^2}(x)dx\\
&&+\int^{\infty}_{0}\log(1+\rho xa_M)f_{|h_M|^2}(x)dx.
\end{eqnarray}
 Even with the approximations in \eqref{cdf11} and \eqref{pdf1}, an exact expression for the ergodic sum rate is still difficult to obtain, and we will focus on the high SNR approximation as well as the asymptotic behavior of the sum rate when $M\rightarrow \infty$.
\setcounter{subsubsection}{0}
\subsubsection{High SNR approximation}
When $\rho\rightarrow \infty$,   the ergodic  sum rate can be expressed as follows:
\begin{align}
R_{ave} &\approx \sum^{M-1}_{m=1}\int^{\infty}_{0}\log \left(1 + \frac{ a_m}{\tilde{a}_m} \right)  f_{|h_m|^2}(x)dx\\\nonumber & +\underset{T_1}{\underbrace{\int^{\infty}_{0}\log \left(1 + x\rho a_M \right)  f_{|h_M|^2}(x)dx}}.
\end{align}
The term $T_1$ can be rewritten  as follows:
\begin{align}
T_1=\frac{\rho a_M}{\ln 2}\int^{\infty}_{0}\frac{1-F_{|h_M|^2}(x)}{1+x\rho a_M}dx.
\end{align}
Rewrite the CDF in the following form:
\begin{eqnarray}\label{CDF 3}
F_{|\tilde{h}|^2}(x)= \frac{1}{\mathcal{R}_D}\sum_{n=0}^{N}b_ne^{-c_nx},
\end{eqnarray}
where   $b_n=-w_n\sqrt{1-\theta_n^2}\left(\frac{\mathcal{R}_D}{2}\theta_n+\frac{\mathcal{R}_D}{2}\right)$ for $1\leq n \leq M$, $b_0=-\sum^{N}_{n=1}b_n$, and $c_0=0$. As a result, the CDF of the largest order statistics is $
F_{| {h}_M|^2}(x)=  \left(F_{|\tilde{h}|^2}(x)\right)^M$.
Now $T_1$ can be expressed as follows:
\begin{align}
T_1&=\frac{\rho a_M}{\ln 2}\int^{\infty}_{0}\frac{1}{1+x\rho a_M}\left(1- \frac{1}{\mathcal{R}_D^M}\underset{k_0+\cdots+k_N=M}{\sum}\right. \\\nonumber &  \left.{M \choose k_0, \cdots,k_N}\left(\prod^{N}_{n=0} b_n^{k_n}\right)e^{-\sum_{n=0}^{N}k_nc_nx}\right) dx,
\end{align}
where ${M \choose k_0, \cdots,k_N}=\frac{M!}{k_0!\cdots k_N!}$. Clearly  the integral $\int^{\infty}_{0}\frac{\epsilon}{1+x\rho a_M} dx$  does not exist, where  $\epsilon$ is a constant. One can  first make the following observation:
\begin{eqnarray}
 \frac{ {M \choose k_0, \cdots,k_N}\left(\prod^{N}_{n=0} b_n^{k_n}\right)e^{-\sum_{n=0}^{N}k_nc_nx} }{\mathcal{R}_D^M} =1,
\end{eqnarray}
when $k_0=N$, and $k_i=0$, $1\leq i \leq N$, since $F_{|\tilde{h}|^2}(\infty)=1$.
This observation can be used to remove the constants in the integral, and $T_1$ is written as follows:
\begin{align}\nonumber
T_1&=- \frac{ \rho a_M}{\mathcal{R}_D^M\ln 2}\int^{\infty}_{0}\frac{1}{1+x\rho a_M} \underset{\textrm{\parbox{.7in}{\vspace{4pt} $k_0+\cdots+k_N=M, k_0\neq M$}}}{\sum}{M \choose k_0, \cdots,k_N}  \\ &\times  \left(\prod^{N}_{n=0} b_n^{k_n}\right)e^{-x\sum_{n=0}^{N}k_nc_n}  dx.
\end{align}
In the above equation, each term of the sum is an exponential function with a non-zero exponent, i.e. $\sum_{n=0}^{N}k_nc_n\neq 0$. With some algebraic manipulations, the ergodic sum rate achieved by NOMA can be obtained as follows:
\begin{align}\label{ergodic rate ss}
R_{ave} &\approx \sum^{M-1}_{m=1} \log \left(1 + \frac{
a_m}{\tilde{a}_m} \right)    - \frac{1}{\mathcal{R}_D^M\ln
2}\\\nonumber &\times \underset{\textrm{\parbox{.7in}{\vspace{4pt}
$k_0+\cdots+k_N=M, k_0\neq M$}}}{\sum}{M \choose k_0, \cdots,k_N}
\left(\prod^{N}_{n=0}
b_n^{k_n}\right)e^{\frac{\sum_{n=0}^{N}k_nc_n}{2\rho
a_M}}\\\nonumber &\times \left(\frac{\sum_{n=0}^{N}k_nc_n}{\rho
a_M}\right)^{-\frac{1}{2}}W_{-\frac{1}{2},0}\left(\frac{\sum_{n=0}^{N}k_nc_n}{\rho
a_M}\right),
\end{align}
where $W_{k,u}(\cdot)$ denotes the Whittaker function.

\subsubsection{Asymptotic study with $M\rightarrow \infty$}
In this subsection, we focus on  the asymptotic performance of   NOMA when $M\rightarrow\infty$.  First define the growth function as $G(x)\triangleq \frac{1-F_{|\tilde{h}|^2}(x)}{f_{|\tilde{h}|^2}(x)}$. Note that  the condition needed in order  to apply the extreme value theorem is that the limit, $\underset{x\rightarrow\infty}{\lim}G(x)$, exists. This condition holds  for the addressed distribution  as shown in the following:
\begin{align}
\underset{x\rightarrow\infty}{\lim}G(x) &= \underset{x\rightarrow\infty}{\lim} \frac{1-\frac{1}{\mathcal{R}_D}\sum^{N}_{n=0}b_ne^{-c_nx}}{\frac{1}{\mathcal{R}_D}\sum^{N}_{n=1}\beta_ne^{-c_nx}}\underset{a}{=}   \frac{-b_N}{\beta_N}  ,
\end{align}
where the step (a) is obtained because  $\frac{1}{\mathcal{R}_D}b_0e^{-c_0x}=1$ and $c_N\leq c_i$ for all $1\leq i\leq N$.

The evaluation of the   asymptotic behavior of $|h_M|^2$ needs $u_M$, the unique solution of $1-F_{|\tilde{h}|^2}(u_M)=\frac{1}{M}$.
This equation can be first rewritten as follows:
\begin{align}\label{app1}
-\frac{1}{\mathcal{R}_D}\sum^{N}_{n=1}b_ne^{-c_nu_M}=\frac{1}{M}.
\end{align}
When $u_M\rightarrow \infty$, \eqref{app1} can be approximated as follows:
\begin{align}
-\frac{b_Ne^{-c_Nu_M}}{\mathcal{R}_D}\left(1+O\left(\frac{1}{u_M}\right)\right)=\frac{1}{M}.
\end{align}
It is worth  pointing  out that the terms $e^{-c_nu_M}$, $m<N$, are decreasing at a   rate faster than $\frac{1}{u_M}$, but the use of the above expression can ensure that  the existing results in \cite{Sharif05} and \cite{David03}  can be applied straightforwardly. Particularly, following steps similar to those used in \cite{Sharif05}, the  the solution $u_M$ is given by
\begin{align}
u_M = \frac{1}{c_N} \log(M) +O(\log\log M) .
\end{align}
Similarly we observe $G^{(m)}(u_M)=O\left(\frac{1}{u_M^m}\right)$. By applying Corollary A1 in \cite{Sharif05}, it is straightforward to show that $
\mathrm{P}\left(\log M -c_N\log\log M \leq |h_M|^2\leq   \log +c_N\log \log M\right) \geq 1-O\left(\frac{1}{\log}\right)$, where $\frac{-b_N}{\beta_N}=c_N$.

Therefore we can conclude that NOMA can achieve the following ergodic sum rate:
\begin{align}\label{asymptodic}
R_{ave} \rightarrow \log ( \rho \log\log M),
\end{align}
with a probability approaching one when $M\rightarrow \infty$ and $\rho \rightarrow \infty$. Consider an opportunistic MA approach that  allocates all the bandwidth resource to the user with the best channel condition. It is easy to verify that this opportunistic scheme achieves the upper bound of the system throughput with an asymptotic behavior of $\log ( \rho \log\log M)$. Therefore NOMA can achieve the same asymptotic performance as the opportunistic scheme, but NOMA can offer better   fairness since all the users are served  simultaneously.
\vspace{-1.2em}
\begin{figure}[!htp]
\begin{center} \subfigure[$\tilde{R}_1=0.1$ BPCU and $\tilde{R}_2=0.5$ BPCU ]{\includegraphics[width=0.37\textwidth]{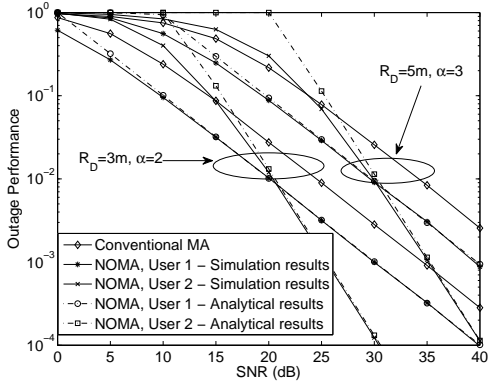}}
 \subfigure[Different   $R_m$ with $\mathcal{R}_D=5m$ and
$\alpha=3$ ]{ \includegraphics[width=0.36\textwidth]{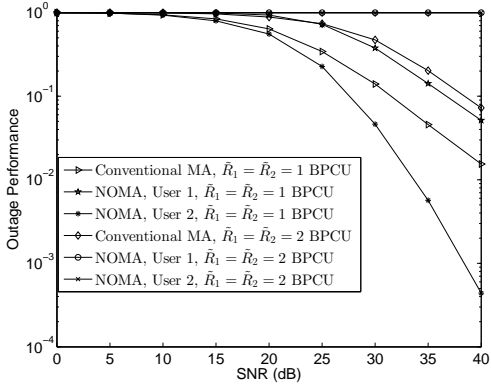}}
\end{center}
\vspace*{-1.5em} \caption{\vspace{-2em}Outage performance of the multiple access technologies   }\label{fig outage}\vspace{-0.5em}
\end{figure}
\vspace{-1.2em}
\section{Numerical Results}\label{section numerical}
\begin{figure}[!htp]
\begin{center} \subfigure[ $M=2$]{\includegraphics[width=0.37\textwidth]{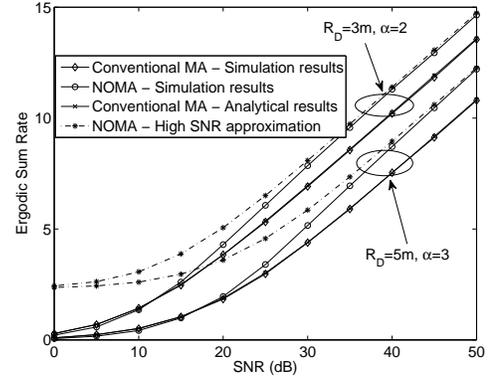}}
\subfigure[Impact of $M$ with $\mathcal{R}=5m$ and $\alpha=2$]{
\includegraphics[width=0.35\textwidth]{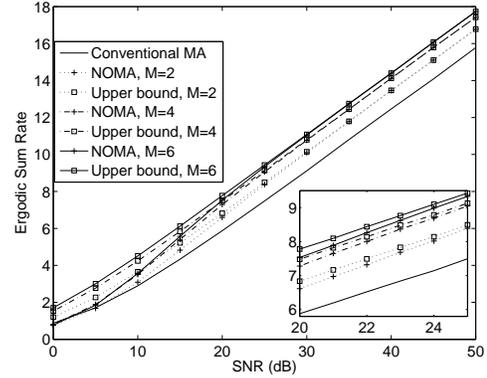}}
\end{center}
\vspace*{-3mm} \caption{ Ergodic sum rates achieved by the multiple access technologies.   }\label{fig ergodic 1}\vspace{-1.8em}
\end{figure}

In this section, the performance of NOMA is evaluated by using computer simulations, where a conventional orthogonal MA approach with {\it a randomly scheduled user} is used for beanchmarking. The power allocation coefficients are $a_1=\frac{4}{5}$ and $a_2=1-a_1$ for $M=2$. For $M>2$, $a_{m}=\frac{M-m+1}{\mu}$ and $\mu$ is to ensure $\sum^{M}_{m=1}a_m=1$. $N=10$. In Fig. \ref{fig outage} the outage performance is shown as a function of SNR, where the targeted rate for the conventional scheme is $ \sum_{m=1}^{M}\tilde{R}_m $ bit per channel use (BPCU). As can be observed from Fig. \ref{fig outage}.a, NOMA outperforms the comparable scheme, and the diversity order of the users is a function of their channel conditions, which is consistent to \eqref{outage 2}. However, with an incorrect  choice of   $\tilde{R}_j$ and $a_m$, the outage probability will  be always one, as shown in Fig. \ref{fig outage}.b. In Fig. \ref{fig ergodic 1}, the ergodic sum rate achieved by NOMA is shown as a function of SNR. The two  figures in Fig. \ref{fig ergodic 1} demonstrate that NOMA can achieve a larger sum rate than the orthogonal MA scheme, and approach the upper bound of the system throughput which is achieved by the opportunistic MA scheme. It is worth pointing  out that the provided simulation results shown in Fig. \ref{fig outage}.a and Fig. \ref{fig ergodic 1}.a match the  analytical results developed at \eqref{outage 2} and \eqref{ergodic rate ss}.

\section{Conclusions}
 In this paper the performance of NOMA has been investigated by using two metrics, the outage probability and ergodic sum rates. We first  demonstrated that  NOMA can achieve better outage performance than the orthogonal MA techniques, under the condition that the users' rates and power coefficients are carefully chosen. In addition, we have shown that NOMA can achieve a superior ergodic sum rate, and is asymptotically equivalent to the opportunistic MA technique. However, there are two potential drawbacks to NOMA. One is that NOMA introduces additional  complexity due to the use of SIC, and the other is that the performance gain of NOMA at low SNR is insignificant. Therefore it is important  to study how to achieve a tradeoff of performance and complexity at different SNRs.
\vspace{-1em}
 \bibliographystyle{IEEEtran}
\bibliography{IEEEfull,trasfer}

\end{document}